\def \SAIT #1 #2 {{\em Mem.\ Soc.\ Astron.\ It.\/} {\bf #1}, #2}
\def \MESS #1 #2 {{\em The Messenger\/} {\bf #1}, #2}
\def \ASTRNACH #1 #2 {{\em Astron. Nach.\/} {\bf #1}, #2}
\def \AAP #1 #2 {{\em Astron. Astrophys.\/} {\bf #1}, #2}
\def \AAL #1 #2 {{\em Astron. Astrophys. Lett.\/} {\bf #1}, L#2}
\def \AAR #1 #2 {{\em Astron. Astrophys. Rev.\/} {\bf #1}, #2}
\def \AAS #1 #2 {{\em Astron. Astrophys. Suppl. Ser.\/} {\bf #1}, #2}
\def \AJ #1 #2 {{\em Astron. J.\/} {\bf #1}, #2}
\def \ANNREV #1 #2 {{\em Ann. Rev. Astron. Astrophys.\/} {\bf #1}, #2}
\def \APJ #1 #2 {{\em Astrophys. J.\/} {\bf #1}, #2}
\def \APJL #1 #2 {{\em Astrophys. J. Lett.\/} {\bf #1}, L#2}
\def \APJS #1 #2 {{\em Astrophys. J. Suppl.\/} {\bf #1}, #2}
\def \APSS #1 #2 {{\em Astrophys. Space Sci.\/} {\bf #1}, #2}
\def \ASR #1 #2 {{\em Adv. Space Res.\/} {\bf #1}, #2}
\def \BAIC #1 #2 {{\em Bull. Astron. Inst. Czechosl.\/} {\bf #1}, #2}
\def \JSQRT #1 #2 {{\em J. Quant. Spectrosc. Radiat. Transfer\/} {\bf #1}, #2}
\def \MN #1 #2 {{\em Mon. Not. R. Astr. Soc.\/} {\bf #1}, #2}
\def \MEM #1 #2 {{\em Mem. R. Astr. Soc.\/} {\bf #1}, #2}
\def \PLR #1 #2 {{\em Phys. Lett. Rev.\/} {\bf #1}, #2}
\def \PASJ #1 #2 {{\em Publ. Astron. Soc. Japan\/} {\bf #1}, #2}
\def \PASP #1 #2 {{\em Publ. Astr. Soc. Pacific\/} {\bf #1}, #2}
\def \NAT #1 #2 {{\em Nature\/} {\bf #1}, #2}

\documentstyle{memsait}
\input epsf.sty
\begin{opening}
\title{ X RAY PRECURSORS IN SGRs: PRECESSING $\gamma$ JET TAILS }
\author{DANIELE FARGION}
\institute{Physics Department and INFN, Rome University 1 -
Ple.A.Moro 2, 00185, Rome, Italy}
\date{} 
\end{opening}

\begin{document}

\oddpagefooter{}{}{} 
\evenpagefooter{}{}{} 
\
\bigskip

\begin{abstract}
 Weak isolated X-ray precursor events before the main Gamma Ray Burst, GRB,
and also rare Soft Gamma Repeaters, SGR, events are in complete
disagreement with any Fireball, or Magnetar,  one-shoot explosive
scenarios. Fireball model in last two years has been deeply
modified into a fountain beamed Jet exploding  and  interacting on
external  shells to explain GRB fine time structure. On the
contrary  earlier we  proposed a unified scenario for both
GRBs-SGRs where a precessing Gamma  Jet (of different intensity)
and its geometrical beaming is the source of both GRB and SGRs
wide morphology.  GRBs are peaked SNs Jet spinning and precessing
observed along the thin Jet axis. Their mysterious weak X
precursors bursts, corresponding to non-negligible energy powers,
up to million Supernova ones for GRB, are $\gamma$ Jet tails
beamed off-axis, observed at X-Ray tails. They are rare, about
$(3-6)\%$ of all GRBs, but not unique at all. Comparable brief
X-ray precursor flashes occurred in rarest and most detailed SGRs
events as the 27 and the 29 August 1998 event from SGR 1900+14.
The same source has been  in very power-full  activity on recent
18 April 2001 once again preceded by X-Ray precursors. These
events are inconsistent with any Fireball or
Magnetar-Mini-Fireball models. We interpret them naturally as
earlier marginal blazing of outlying X conical precessing Jet, an
off-axis tails surrounding a narrower gamma precessing Jet. Only
when the light-house Jet is in on-axis blazing mode toward the
Earth we observe the harder power-full SGR event. We predict such
a rich X-Ray precursor signals (more numerous then gamma ones)
during Soft Gamma Repeater peak activities; they should be
abundant and  within detection threshold by a permanent
monitoring SGRs by Beppo-Sax WFC or Chandra X ray satellites
while at peak activity.
\end{abstract}

\section{Introduction: GRBs and SGRs: Spinning, Precessing and  Blazing $\gamma$ Jet}

Gamma Ray Burst and Soft Gamma Repeaters reached an apparent stage
of maturity: tens of GRBs found, finally, an X, optical and (or)
radio transient (the after-glow) identification as well as some
associated host galaxies at cosmic red-shifts (Bloom et all 2000).
New categories of GRBs and SGRs events have been labeled, but
even within these wider updated data no conclusive theory or even
partial understanding seem to solve the old-standing GRB/SGRs
puzzle: the nature of The GRB-SGR signals.
  On the contrary the wider and wider collection of data are
  leading to a schizophrenic attitude in the most popular
  isotropic models, the Largest Cosmic Explosions (Fire-ball, Hypernova,
  Supra-Nova) with more and more phenomenological
  descriptions (power laws everywhere) and less and less unifying
  views. This "give up" attitude seem to reflect the surprising
  never ending morphologies of GRBs.
  We argued on the contrary since  1994 and in particular after
  1998  that GRBs and SGRs find a comprehensive theory within a thin spinning
  and multi precessing $\gamma$ Jet, sprayed by a Neutron Star, NS, or Black Hole,
  BH; the gamma jet is fed by electron pair GeV Jet in Inverse
  Compton Scattering. (Fargion 1994-1999, Fargion, Salis 1995-1998).
  Isotropic Fireball models , which enjoyed a decade of total
  predominance, with the  extreme GRB energy released  ($\gg$ $10^{54}$
  $erg $), comparable to few solar masses annihilation or more, lead to a deep
   conflict with any  energy and masses. Indeed their  corresponding Schwarchild
   scale times are above milliseconds and disagree with the observed sharp
   GRBs fine time structures  much below a fraction of millisecond.
    Fireball Isotropic models simply  ignored the role
   of thermal neutrinos which imply at least twice the observed   $\gamma$  energy
 in GRBs. Fireball-Jet (a crazy acronym like the spherical-knife) mitigate the energy
 puzzle but cannot explain the huge wide energy range among known
 GRBs: indeed the energy power spread (from $10^{53}$
  $erg s^{-1}$) for most far GRBs versus $10^{46}$  $erg  s^{-1}$ for
  nearest GRB980425, led most Fireball defenders just
   to neglect, hide or even reject in a very arbitrary (and partisan) way
  the  nearest and best identified GRB connection to a Supernova, SN,
  explosion as SN1998bw.  We naturally interpreted this event as a slightly
   off-axis blazing Jet.  We also understand the same rarity of GRB-SN detection
  and the established GRB980425-SN link as an additional
  evidence for the thin Jet nature of GRBs.
   Even originally ($1970-80$) unified  GRB/SGR models since last fifteen years
   are commonly separated by their repeater and spectra differences;
 however  very recently they openly shared
  the same spectra, time and flux structures (Fargion 1998-99,Woods et all 1999).
  This analogy suggest a common nature.
   Their different distances, cosmic versus galactic ones, imply
different power source Jet. The $\gamma$ Jet is born by high GeVs
electron pairs Jet which are regenerating, via Inverse Compton
Scattering, an inner collimated beamed $\gamma$ (MeVs)-(tens
KeVs) X precessing  Jet. The thin jet (an opening angle inverse of
the electron Lorentz factor, nearly milli-radiant), while
spinning, is driven by a companion and/or an asymmetric accreting
disk in a Quasi Periodic Oscillation (QPO) and in a Keplerian
multi-precessing blazing mode: its $\gamma-X$ ray lighthouse
trembling and flashing is the source of the complex and wide
structure of observed Gamma Bursts.  These $\gamma$ Jets share a
peak power of a Supernova ($10^{44} erg s^{-1}$) at their birth
(during SN and Neutron Star formations), decaying by power law
$\sim t^{-1}$ $-$ $\sim t^{-(1.5)}$ to less power-full Jets that
converge to present persistent SGRs stages. Indeed SGRs  are
blazing events from late relic X pulsar Jet with no associated
explosion (or OT and afterglow) observable only at nearer
distances in axis. The $\gamma$ Jet responsable for SGR emit in
general at $\sim$ $ 10^{35}$ erg$ s^{-1}$ Jet powers comparable to
the angular momentum losses; it is the thin angular collimation
and not the Magnetar field budget, to lead to peak apparent SN
powers $\sim$ ($10^{44}$) erg$ s^{-1}$ . In analogy GRB  show an
apparent luminosity  of a SN $\sim$ ($10^{44}$) erg$ s^{-1}$
amplified by the inverse square of the thin angle, from $10^{-3}$
to $10^{-4}$ radiant angle Jet beaming : the corresponding solid
angle $\Omega$ spreads between $10^{-7}$ and $10^{-9}$ and the
apparent amplification enjoy a huge factor:$10^{7}$ up to
$10^{9}$ corresponding to GRBs .
 Optical-Radio After-Glows are not only the fading Supernovae
explosion tails often observed in puzzling variable non monotonic
decay, but they exhibith the averaged external Jet beamed tails
moving and precessing while fading away. The rare optical
re-brightening (the so called SN bump) observed in few afterglow
might be erroneously associated to an underlying isotropic SN
flash: it is probably the late re-crossing of the precessing Jet
periphery toward the observer direction. The integral GRB-Jet
activity may leave a trace  triggering giant arc star formations.


A convincing evidence against any explosive GRB model, confirming
present $\gamma$ precessing Jet theory, is hidden in the in the
recent GRB 000131 data which show an un-explicable, for Fireball
model, X precursor signal 7 sec long, just ${62}$ seconds before
to the huge main gamma trigger. No  Fireball or Hypernova
progenitor (binary NS or BH) could survive such a disruptive X
ray precursor power $\gg 10^{48} erg sec^{-1}$.
\\

\begin{figure}[thb]
  \epsfxsize= 6 cm 
  \epsfbox{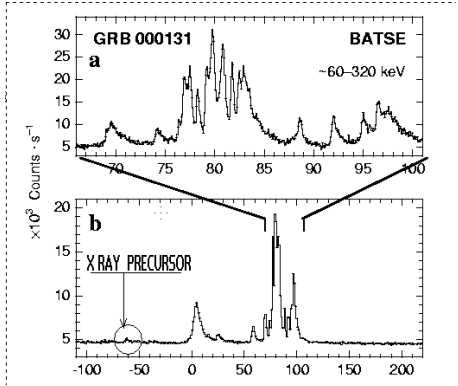}  
  \caption{\em {Location and Intensity of early $X$ Precursor in GRB000131}}
\end{figure}

\begin{figure}[thb]
\mbox{
\epsfxsize=.4\textwidth
 \epsfbox{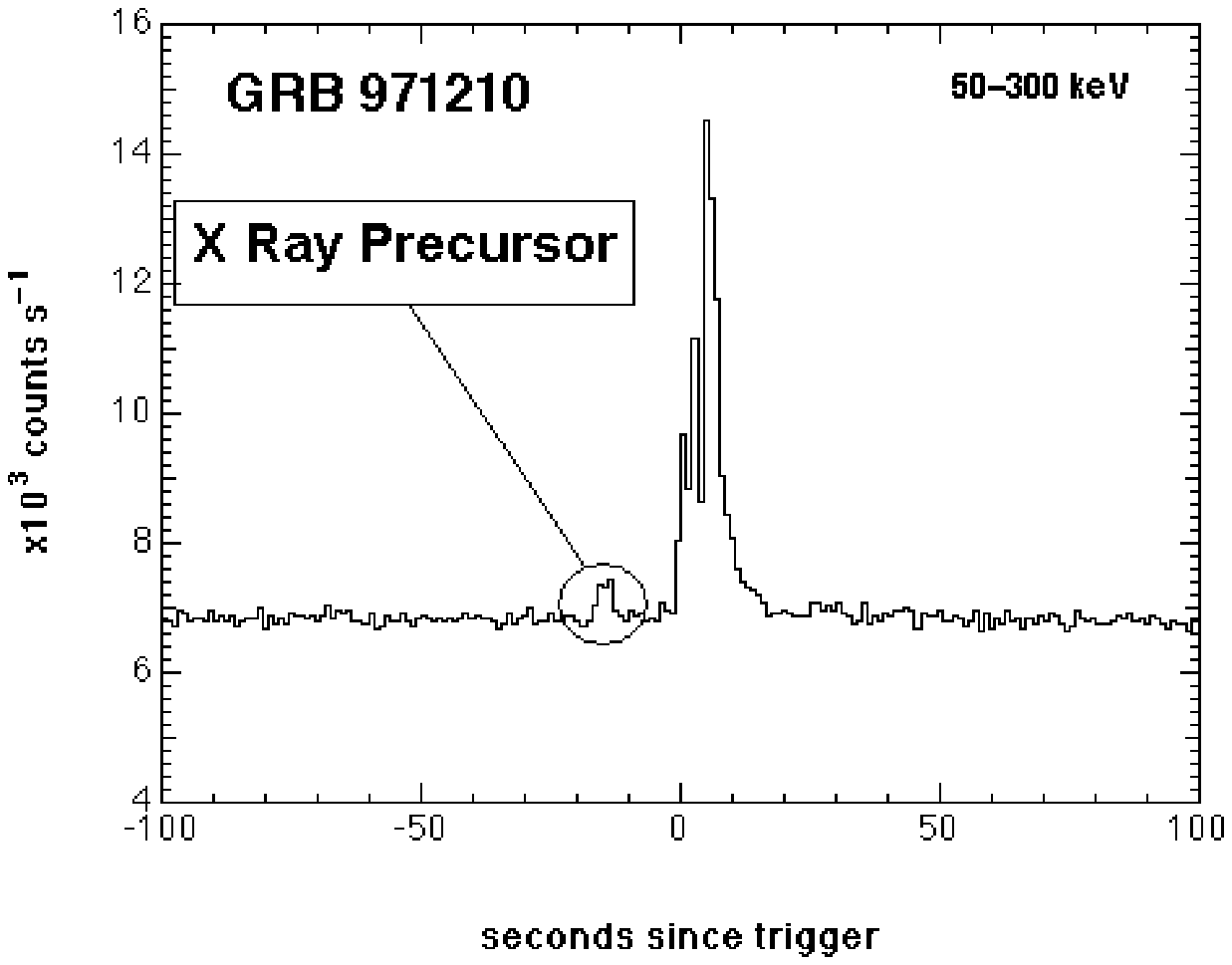}  
\epsfxsize=.5\textwidth 
 \epsfbox{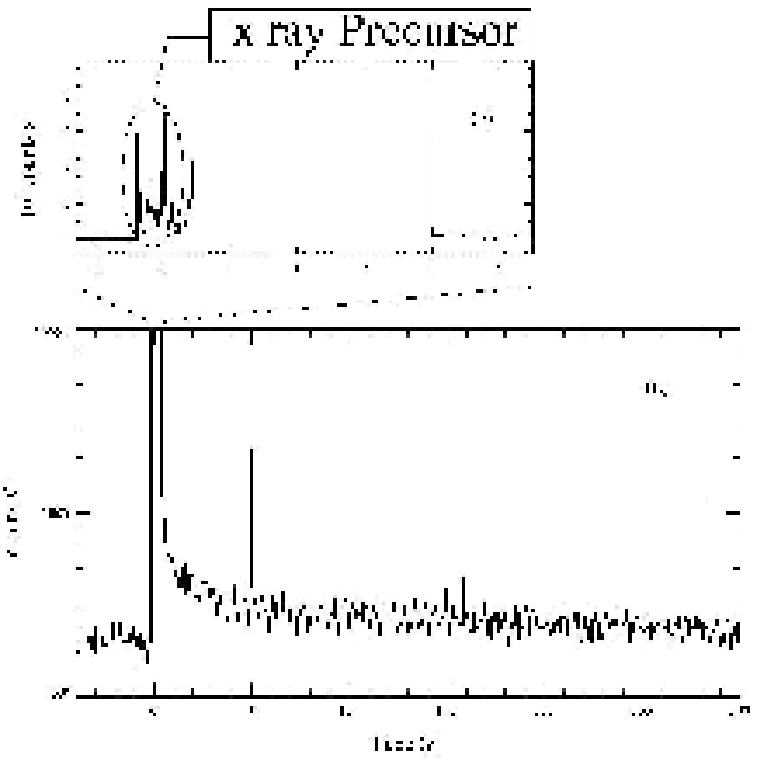} 
}
 \caption{\em {Fig $2a$ and $2b$: Time evolution and X precursors in
 GRB $971210$ and SGR $1900+14$ taking place on 27 August 1998 }}
\end{figure}


\section{ Are GRBs an Unique $\gamma$ Jet Explosion?}

  As we noted above  GRBs are not showing any standard candle
  behaviours within any Fireball isotropic model. Some moderate wide
  Fireball-Jet (a not conclusive compromise, like so called collapsar combining SN
  and open fountain-like Jet explosion) models with a  large beaming (ten degree
  opening) can accommodate all cosmic GRBs excluding the "problematic" nearest
  $GRB980425$ event  keeping it out of the frame.
  Nevertheless SGRs, which share, some time, the same GRB
  signature, are themselves still within a popular isotropic
  mini-Fireball scenario powered by Magnetar  explosive
  events. This model imply a magnetic energy in the neutron star at least 4
  order of magnitude above the kinetic rotational energy, calling for
  an anomalous and unexplained energy equipartition bias.
  Beaming may solve the puzzle within a common X-ray Pulsar power.

    Indeed  the SGR1900+14 event BATSE   trigger 7171 left an almost
     identical event comparable to a just following  GRB (trigger 7172) on
   the same day, same detector, with same spectra and comparable flux.
      This Hard-Soft connection has been re-discovered
      and confirmed   more recently  by BATSE group: ( Fargion 1998-1999;Woods et all
      1999) with an additional hard event of SGR 1900+14 recorded in GRB990110 event.
    An additional GRB-SGR connection occur between  GRB980706
   event with an almost identical (in time, channel spectra, morphology and
   intensities) observed in GRB980618 originated by SGR 1627-41.
    Nature would be   very perverse in mimic two signals,
     (even  if at different distances and different powers),
    by two totally  different source engines.The jet angle is related
     by a relativistic kinematics
$\theta \sim \frac{1}{\gamma_e}$, where $\gamma_e$ is found to
reach $\gamma_e \simeq 10^3 \div 10^4$ (Fargion 1994, 1998). At
first approximation the gamma constrains is given by Inverse
Compton relation: $< \epsilon_\gamma > \simeq \gamma_e^2 \, k T$
for $kT \simeq 10^{-3}-10^{-1}\, eV$ and $E_e \sim GeVs$ leading
to characteristic X-$\gamma$ GRB spectra.  The origin of $GeVs$
electron pairs are  very probably decayed secondary related to
primary inner muon pairs jets, able to cross dense stellar
target. The a-dimensional photon number rate (Fargion \& Salis
1996) as a function of the observational angle $\theta_1$
responsible for peak luminosity  has been found:
\begin{equation}
\frac{\left( \frac{dN_{1}}{dt_{1}\, d\theta _{1}}\right) _{\theta
_{1}(t)}}{ \left( \frac{dN_{1}}{dt_{1}\, d\theta _{1}}\right)
_{\theta _{1}=0}}\simeq \frac{1+\gamma ^{4}\, \theta
_{1}^{4}(t)}{[1+\gamma ^{2}\, \theta _{1}^{2}(t)]^{4}}\, \theta
_{1}\approx \frac{1}{(\theta _{1})^{3}} \;\;.\label{eq4}
\end{equation}
The total fluence at minimal impact angle $\theta_{1 m}$
responsible for the average luminosity  is
\begin{equation}
\frac{dN_{1}}{dt_{1}}(\theta _{1m})\simeq \int_{\theta
_{1m}}^{\infty }\frac{ 1+\gamma ^{4}\, \theta _{1}^{4}}{[1+\gamma
^{2}\, \theta _{1}^{2}]^{4}} \, \theta _{1}\, d\theta _{1}\simeq
\frac{1}{(\, \theta _{1m})^{2}}\;\;\;. \label{eq5}
\end{equation}
These spectra fit GRBs observed ones (Fargion \& Salis 1995).
Assuming a beam jet intensity $I_1$ comparable with maximal SN
luminosity, $I_1 \simeq 10^{45}\;erg\, s^{-1}$, and replacing
this value in adimensional expressions above we find a maximal
apparent GRB power for beaming angles $10^{-3} \div 3\times
10^{-5}$, $P \simeq 4 \pi I_1 \theta^{-2} \simeq 10^{52} \div
10^{55} erg \, s^{-1}$ within observed ones. We also assume a
power law jet time decay as follows
\begin{equation}\label{eq6}
  I_{jet} = I_1 \left(\frac{t}{t_0} \right)^{-\alpha} \simeq
  10^{45} \left(\frac{t}{3 \cdot 10^4 s} \right)^{-1} \; erg \,
  s^{-1}
\end{equation}
where ($\alpha \simeq 1$ or $\alpha \simeq 1.5$) able to reach, at
1000 years time scales, the present known galactic microjet (as
SS433) intensities powers: $I_{jet} \simeq 10^{39}-10^{36}\;erg\,
s^{-1}$. We used the model to evaluate if April 1998 precessing
jet might hit us once again (as it possibly did on GRB980712) . It
should be noted that a steady angular velocity would imply
(during the earliest few hours) an average intensity variability
($I \sim \theta^{-2} \sim t^{-2}$) corresponding to some of the
earliest afterglow decay law. At later stages a rare re-beaming
of the precessing Jet combined with a  power Jet decay may alter
the afterglow decay power law. Such a  Jet  blazing beaming and
re-hitting may be the source of rarest re-brightening observed on
GRB980508 and on GRB000301C after-glows.

\section{X precursors in  SGRs  by Precessing Jet}

We imagine the precessing Jet nature as the late stages of jets
fueled by a disk or a companion (WD, NS) star. Their binary
angular velocity $\omega_b$ reflects the beam evolution
$\theta_1(t) = \sqrt{\theta_{1 m}^2 + (\omega_b t)^2}$ or more
generally a multi-precessing angle $\theta_1(t)$ (Fargion \&
Salis 1996):

\begin{equation}\label{eq7}
  \theta_1(t) = \sqrt{\theta_{x}^2 +\theta_{y}^2 }
\end{equation}

\begin{equation}\label{eq8}
  \theta_{x}(t) =                               
  \theta_{b} sin(\omega_{b} t + \varphi_{b})+
  \theta_{psr}sin(\omega_{psr} t)+
  \theta_{N}sin(\omega_{N} t  + \varphi_{N})
\end{equation}

\begin{equation}\label{eq8}
  \theta_{y}(t) = \theta_{1 m}+
  \theta_{b} cos(\omega_{b} t + \varphi_{b})+
  \theta_{psr} cos(\omega_{psr} t)+
  \theta_{N} cos(\omega_{N} t  + \varphi_{N})
\end{equation}
where $\theta_{1 m}$ is the minimal angle impact parameter of the
jet toward the observer, $\theta_{b}$, $\theta_{psr}$,
$\theta_{N}$ are, in the order, the maximal opening precessing
angles due to the binary, spinning pulsar, nutation mode of the
jet axis. Additional multi precessions are also possible leading
to more complex and realistic $\gamma$ burst evolution.

The angular velocities combined in the multi-precession keep
memory of the pulsar jet spin ($\omega_{psr}$), the precession by
the binary $\omega_b$ and an additional nutation due to inertial
momentum anisotropies or beam-accretion disk torques ($\omega_N$).

 The relativistic morphology of the Jet and its multi-precession is
the source of the puzzling complex $X$-$\gamma$ spectra signature
of GRBs and SGRs. Its inner internal Jet contain, following the
relativistic Inverse Compton Scattering,  hardest and rarest
beamed GeVs-MeVs photons (as the rarest EGRET GRB940217 one) but
its external Jet cones are dressed by softer and softer photons.
This   onion like multi Jets is not totally axis symmetric: it
doesn't appear on front as a concentric ring serial; while
turning and spraying around it is deformed (often) into an
elliptical off-axis concentric rings preceded by the internal
Harder center leading to a common Hard to Soft GRBs (and SGRs)
train signal. In our present model and simulation this internal
effect has been  here neglected without any major consequence. The
complex variability of GRBs and SGRs are simulated successfully
by the equations and the consequent geometrical beamed Jet
blazing leading to the observed $X-\gamma$ signatures. As shown
in Fig. 3 the slightly different precessing configurations could
easely mimic the wide morphology of GRBs and SGRs as well as the
surprising rare X-ray precursor observed in Fig.1.2 and simulated
below for an event comparable to SGR1900+14 on August 1998. In
conclusion the X-Ray precursor must occur often around main SGR
$\gamma$ event. We predict that such precessing Jet imprint must
be soon found by continous monitoring of active SGRs. The very
recent discover of such X-Ray precursors (Feroci M. et all, GCN
1060), at 2537, 755, and 444 s before the giant flare on 18
April  with durations of 100, 125 and 55 ms, give additional
support to our model predictions. The author thanks Pier Giorgio
De Sanctis Lucentini and Andrea Aiello for kind support and
numerical test.

\begin{figure}[thb]
  \epsfxsize=1.0\textwidth 
  \epsfbox{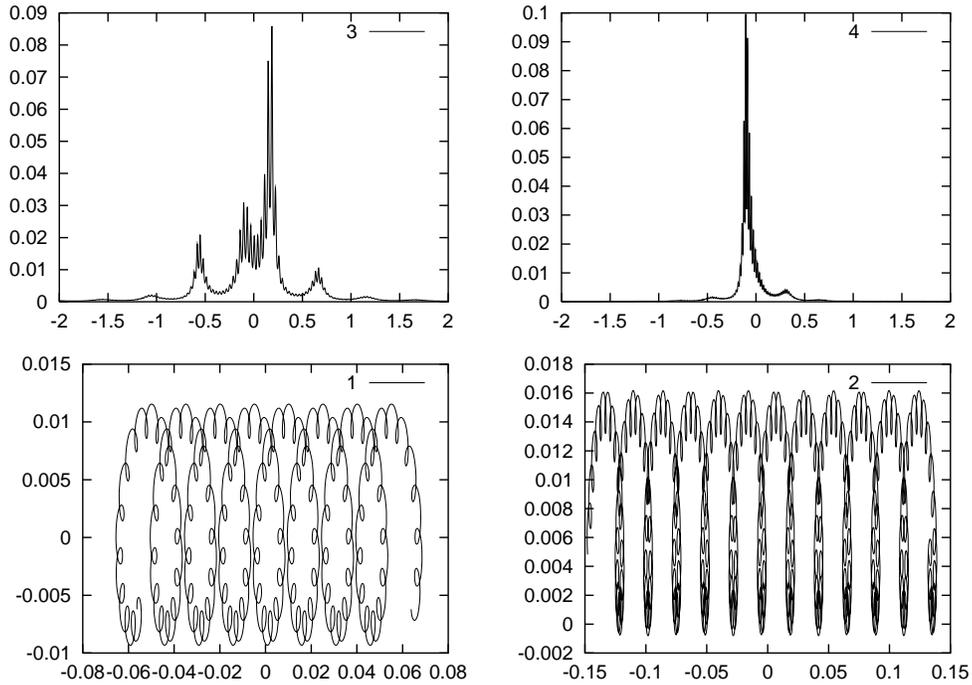}  
  \caption{\em{ Label 1 and 2 in Figure  below show two different
  bi-dimensional angle Spinning, Precessing Gamma Jet ring patterns toward the
   detector at the  origin ($0,0$).
   The corresponding  Label 3-4 in Figure above show the consequent X,
    $\gamma$ flux intensity time evolution
   derived by the ICS formula in the text. X ray precursors,
    as the events on August 1998  or on April 2001 from SGR1900+14  could be related to similar ideal or more complicated
    precessing beam Jet patters. }}
\end{figure}

\end{document}